\documentclass[aps,prd,reprint,superscriptaddress,nofootinbib,floatfix]{revtex4-2}

\usepackage{graphicx}
\usepackage{amsmath}
\usepackage{amsfonts}
\usepackage{amssymb}
\usepackage{xcolor}
\usepackage{times,txfonts}
\usepackage{mathtools}
\usepackage[hidelinks,pdfusetitle,hypertexnames=false]{hyperref}
\usepackage{ulem}

\definecolor{coauthorgreen}{RGB}{0,128,0}

\begin{document}


\title{Corrections to the Unruh Effect from Robin Boundary Conditions in Punctured Minkowski Spacetime}


\author{Nickolas P. Botta}
\email{n183507@dac.unicamp.br}
\affiliation{Instituto de F\'isica Gleb Wataghin, Universidade Estadual de Campinas, 13083-859 Campinas, S\~ao Paulo, Brazil}

\author{Jo\~ao Paulo  M.  Pitelli}
\email[]{pitelli@unicamp.br}
\affiliation{Departamento de Matem\'atica Aplicada, Universidade Estadual de Campinas,
13083-859 Campinas, S\~ao Paulo, Brazil}

\author{Ricardo A. Mosna}
\email{mosna@unicamp.br}
\affiliation{Instituto de F\'isica Gleb Wataghin, Universidade Estadual de Campinas, 13083-859 Campinas, S\~ao Paulo, Brazil}

\begin{abstract}
We consider a real massless scalar field in punctured Minkowski spacetime, endowed at the removed origin with the stable one-parameter family of Robin boundary conditions $G(0)-\beta G'(0)=0$, $\beta\geq0$. We obtain the boundary-induced part of the static ground-state Wightman function in closed form and study the response of a uniformly accelerated Unruh-DeWitt detector. The puncture breaks the boost symmetry underlying the stationary Unruh response, so the boundary-induced pullback is nonstationary in proper time. The corresponding subtracted detector contribution can either enhance or suppress the ordinary Minkowski response and depends on the Robin parameter, the detector gap, and the portion of the trajectory sampled. For the unshifted, radially aligned trajectory studied in detail, we prove that the boundary-induced pullback is absolutely integrable in the two proper-time variables. Consequently, as the smooth interaction is extended over the full detector history, this contribution approaches a finite limit and remains $O(1)$, producing no additional term linear in the interaction duration. The formal Neumann limit is singular: the boundary-induced two-point function grows logarithmically because of an infrared singularity in the $s$-wave sector, and the numerical response exhibits growth consistent with this asymptotic behavior.
\end{abstract}

\maketitle

\section{Introduction}
\label{intro}

The Unruh effect can be formulated operationally through the stationary response of a uniformly accelerated Unruh-DeWitt (UDW) detector in the Minkowski vacuum. Along a boost orbit of proper acceleration $a$, the pulled-back two-point function depends only on the proper-time difference and satisfies the Kubo-Martin-Schwinger (KMS) condition at the Unruh temperature $T_{\mathrm U}=a/(2\pi)$~\cite{unruh,unruh-wald,crispino}. This stationary description relies on the boost invariance of the Minkowski vacuum. A static boundary or defect may preserve inertial-time translations while breaking precisely that boost symmetry, in which case the accelerated response need not remain stationary even though the vacuum state is static.

Static non-globally hyperbolic spacetimes provide a natural setting for this question. There, the field dynamics is not fixed by Cauchy data alone and must be supplemented by a positive self-adjoint extension of the spatial operator~\cite{waldart1,waldart2}. Punctured Minkowski spacetime, obtained by removing the timelike line $\mathcal L=\{(t,\mathbf 0):t\in\mathbb{R}\}$, is the simplest exactly tractable example. It is not a genuine gravitational singularity: the punctured manifold is a proper subset of ordinary Minkowski spacetime and admits a geodesically complete extension. Nevertheless, its value lies precisely in its simplicity. Only the $s$-wave sector admits a nontrivial Robin family, and the same family is equivalent to an elastic zero-range point interaction with scattering length $a_s=-\beta$~\cite{desena}. Thus the model may also be viewed as the idealized limit of an infinitely heavy pointlike scatterer, provided that dynamical properties of a physical particle are neglected~\cite{albeverio}.

Our central question is how a uniformly accelerated UDW detector coupled to this real massless scalar field responds in the static ground state associated with a chosen Robin extension. Punctured Minkowski spacetime may be regarded as the zero-solid-angle-deficit limit of the idealized global-monopole spacetime, with the origin kept removed. This problem is therefore related to, but distinct from, the stationary thermal-detector analysis of the global-monopole background in Ref.~\cite{pitelli1}. There the detector is held at fixed radius and probes a state that is KMS with respect to the same static time translations preserved by the boundary condition. Here, by contrast, the field is in the static ground state while the detector follows a boost orbit. In complete Minkowski spacetime these descriptions are connected by the Unruh effect, but the puncture breaks the boost symmetry that supports that connection. Punctured Minkowski spacetime therefore isolates the difference between a static thermal response and an accelerated response in a static ground state.

In what follows, we derive the static ground-state Wightman function in closed form and decompose it as $W=W_{\mathrm M}+W_\beta$, where $W_{\mathrm M}$ is the ordinary Minkowski term and $W_\beta$ is induced by the Robin boundary condition. For the radially aligned hyperbola, the boundary-induced term is nonstationary in proper time: although the puncture and the state are static, the puncture breaks the boost symmetry on which the stationary Unruh response relies. The corresponding subtracted detector response depends nontrivially on the Robin parameter $\beta$, the detector gap $\Omega$, and the portion of the trajectory sampled. It can either enhance or suppress the Minkowski response and, within the families of trajectories and switching windows considered here, becomes more pronounced as the detector approaches the puncture.

We further show analytically that, for every fixed finite $\beta$, this pullback is absolutely integrable in the two proper-time variables. As the interaction is extended over the full detector history, dominated convergence then gives a finite $T\to\infty$ limit, while the correction remains $O(1)$ rather than generating an additional $O(T)$ contribution. The formal Neumann limit is qualitatively different: the boundary-induced two-point function $W_\beta$ has a logarithmic infrared divergence associated with the low-frequency $s$-wave sector.

The paper is organized as follows: in Sec.~\ref{sec:field}, we construct the Robin extensions, give their point-interaction interpretation, and define the static ground state. In Sec.~\ref{sec:detector-response}, we develop the accelerated-detector response, analyze proper-time nonstationarity, and prove the infinite-duration limit. In Sec.~\ref{sec:distance}   we study the dependence on the trajectory and switching window. Finally, Sec.~\ref{sec:conclusion} summarizes the results.

\section{Field Dynamics in Punctured Minkowski Spacetime}\label{sec:field}

We consider Minkowski spacetime with line element
\begin{equation}
    ds^2=-dt^2+dr^2+r^2d\Omega^2,
\end{equation}
where \(d\Omega^2\) is the metric on the unit two-sphere. After separating variables as
\begin{equation}
    \Phi(x)=e^{-i\omega t}Y_\ell^m(\theta,\varphi)R_{\ell,m,\omega}(r),\,\,\,\ell=0,1,2,\ldots,\quad -\ell\leq m\leq \ell,
\end{equation}
the radial part of the Klein-Gordon equation
\begin{equation}
    \Box \Phi(x)=0
\end{equation}
becomes
\begin{equation}
    R_{\ell,m,\omega}''(r)+\frac{2}{r}R_{\ell,m,\omega}'(r)+\left[\omega^2-\frac{\ell(\ell+1)}{r^2}\right]R_{\ell,m,\omega}(r)=0.
    \label{radial}
\end{equation}
The solutions to Eq.~(\ref{radial}) are
\begin{equation}
    R_{\ell,m,\omega}(r)=A_{\ell, m, \omega} j_{\ell}(\omega r)+B_{\ell, m, \omega} \eta_{\ell}(\omega r),
\end{equation}
where \(j_\ell\) and \(\eta_\ell\) are the spherical Bessel and Neumann functions, respectively. For \(\ell=0\), the convention used here is \(\eta_0(z)=-\cos z/z\). When working in Minkowski spacetime and imposing regularity at \(r=0\), one usually discards the \(\eta_\ell\) solutions. Indeed, \(\eta_\ell(\omega r)\) is not square-integrable near \(r=0\) for \(\ell \geq 1\). For $\ell=0$, $\eta_0(\omega r)$ is square-integrable near $r=0$. However, applying the wave operator to \(\eta_0\) yields a contribution proportional to a Dirac delta distribution,
\begin{equation}
   (\nabla^2+\omega^2)\eta_0(\omega r)=\frac{4\pi}{\omega}\delta^{(3)}(\mathbf{x}),
    \label{delta}
\end{equation}
and hence \(\eta_0\) does not define a regular solution of the homogeneous field equation.
 In punctured Minkowski spacetime, however, $\mathcal{L}=\{(t,\mathbf{x}=\mathbf{0}):t\in\mathbb{R}\}$ is not part of the spacetime under consideration. The delta contribution in Eq.~\eqref{delta} is localized at the removed point and therefore does not define a source on the punctured manifold. Away from \(r=0\), the \(\eta_0\) term solves the homogeneous equation and must also be included among the admissible modes.

Notice that, for \(\ell=0\), defining \(G(r)=rR_0(r)\) leads to
\begin{equation}
    G''(r)+\omega^2G(r)=0.
    \label{eq:G}
\end{equation}
It is then clear that Eq.~(\ref{eq:G}) requires a boundary condition at \(r=0\). We choose a Robin boundary condition,
\begin{equation}
    G(0)-\beta G'(0)=0.
\end{equation}
This condition renders the spatial part of the wave operator self-adjoint, thereby leading to a physically reasonable dynamics for the field~\cite{waldart1,waldart2}.

\subsection{Point-interaction interpretation}\label{sec:point-interaction}

The Robin family has the standard interpretation of an elastic zero-range point interaction in the \(s\)-wave sector~\cite{desena}. To make the relation explicit, write a reduced radial mode as
\begin{equation}
 G_\omega(r)=A_\omega\sin\!\left[\omega r+\delta_0(\omega)\right].
\end{equation}
The boundary condition then gives
\begin{equation}
 \tan\delta_0(\omega)=\beta\omega,
 \qquad
 \omega\cot\delta_0(\omega)=\frac{1}{\beta}.
 \label{phase-shift}
\end{equation}
As discussed in Ref.~\cite{desena}, the $s$-wave scattering amplitude for a short-range  potential takes the form
\begin{equation}
    f_0(\omega)=\frac{1}{\omega\cot\delta_0(\omega)-i\omega}.
\end{equation}
Comparing Eq.~\eqref{phase-shift} with the effective-range expansion in the nuclear-physics sign convention, 
\begin{equation}
 \omega\cot\delta_0(\omega)=-\frac{1}{a_s}+O(\omega^2),
\end{equation}
we identify
\begin{equation}
 a_s=-\beta.
 \label{scattering-length}
\end{equation}
Thus, \(\beta=0\) corresponds to the ordinary regular Minkowski
extension and produces no \(s\)-wave scattering. For $\beta=-a_s\neq0$, the \(s\)-wave scattering amplitude
is
\begin{equation}
    f_0(\omega)    =
    \frac{1}{-\frac{1}{a_s}-i\omega}.
    \label{eq:scattering-amplitude}
\end{equation}
For \(a_s>0\), this amplitude has a pole at
\(\omega=i/a_s\) on the physical sheet, corresponding to a bound state
with spatial eigenvalue \(-1/a_s^2\). This is
precisely the unstable mode that motivates our restriction to
\(\beta\geq0\). The stable branch corresponds to
\(a_s\leq0\), for which no normalizable bound state is present. The formal limit \(\beta\to\infty\) is the infinite-scattering-length threshold resonance. More generally, at the level of elastic propagation outside \(r=0\), the Robin family is precisely the static zero-range point-scatterer model described in Sec.~\ref{intro}.

\subsection{Mode expansion and static ground state}

The normalized radial modes are
\begin{equation}
    R_{\ell}(r) = \sqrt{\dfrac{\omega}{\pi}}\,\dfrac{\left[j_\ell(\omega r)-\omega\beta_\ell \,\eta_\ell(\omega r)\right]}{\sqrt{1+\beta_\ell^2\,\omega^2}} \hspace{0.2cm},
\end{equation}
with
\begin{equation}
     \beta_\ell =
     \begin{cases}
        \beta\in[0,\infty), &\text{if }\ell=0,\\
        0, &\text{if }\ell>0.
     \end{cases}
\end{equation}
As discussed above,  \(\beta \geq 0\), since negative values give rise to unstable modes, as shown in Refs.~\cite{pitelliunstable1,pitelliunstable2}~\footnote{In particular, Ref.~\cite{pitelliunstable2} analyzes the contribution of the unstable mode alone to the Unruh effect. It is shown there that the corresponding response function takes the form of an unnormalized Breit-Wigner distribution.}. The endpoint \(\beta=0\) gives the Dirichlet condition, whereas the formal limit \(\beta\to\infty\) corresponds to the Neumann condition.

The modes
\begin{equation}
\begin{aligned}
    u_{\ell,m,\omega}(x)&=e^{-i\omega t} Y_\ell^m(\theta,\varphi)R_{\ell}(r),\\
    u^\ast_{\ell,m,\omega}(x)&=e^{i\omega t}Y_\ell^{m\ast}(\theta,\varphi)R_{\ell}(r),
\end{aligned}
\end{equation}
form a complete set. For each fixed finite \(\beta\geq0\), we select \(u_{\ell,m,\omega}\), with \(\omega>0\), as the positive-frequency modes with respect to the static Killing field \(\partial_t\), and define\footnote{Other quasifree states on the same self-adjoint extension would, in general, have different two-point functions and detector responses.}
\begin{equation}
 \hat a_{\ell,m,\omega}|0_\beta\rangle=0.
 \label{static-vacuum}
\end{equation}

With this choice, the field operator has the expansion
\begin{equation}
    \hat\Phi(x)=\sum_{\ell=0}^{\infty}\sum_{m=-\ell}^{\ell}\int_0^{\infty}d\omega\,
    \left(\hat a_{\ell,m,\omega}u_{\ell,m,\omega}(x)+\hat a_{\ell,m,\omega}^\dagger u^\ast_{\ell,m,\omega}(x)\right).
\end{equation}
The Wightman function in this state then reads
\begin{equation}
\begin{aligned}
 W(x,x')&:=\langle0_\beta|\hat\Phi(x)\hat\Phi(x')|0_\beta\rangle\\
 &=\sum_{\ell=0}^{\infty}\sum_{m=-\ell}^{\ell}\int_0^{\infty}d\omega\,
 u_{\ell,m,\omega}(x)u^\ast_{\ell,m,\omega}(x') \\
 &=\sum_{\ell=0}^\infty\sum_{m=-\ell}^\ell \int_0^\infty d\omega \,
 \frac{\omega}{\pi}
 \frac{j_\ell(\omega r)-\omega\beta_\ell\eta_\ell(\omega r)}
 {\sqrt{1+\beta_\ell^2\omega^2}}\nonumber\\
 &\qquad\times
 \frac{j_\ell(\omega r')-\omega\beta_\ell\eta_\ell(\omega r')}
 {\sqrt{1+\beta_\ell^2\omega^2}}
 Y_\ell^m(\theta,\varphi)Y_\ell^{*m}(\theta',\varphi')\\
 &\qquad\times e^{-i\omega(t-t')}.
\end{aligned}
\end{equation}
This expression can be written as
\begin{equation}
    W(x,x') = W_{\mathrm M}(x,x') + W_\beta(x,x'),
\end{equation}
with
\begin{equation}
\begin{aligned}
      W_{\mathrm M}(x,x') &= \sum_{\ell=0}^\infty \sum_{m=-\ell}^{\ell} \int_0^\infty d\omega \,
      \dfrac{\omega}{\pi}\, j_\ell(\omega r)j_\ell(\omega r')\\
      &\qquad\times Y_\ell^m(\theta,\varphi)Y_\ell^{*m}(\theta',\varphi')e^{-i\,\omega(t-t')},
\end{aligned}
\end{equation}
and
\begin{equation}\label{wightmanrobinexpansion}
\begin{aligned}
 W_{\beta}(x,x')&=\frac{1}{4\pi^2}\int_0^\infty d\omega\,\omega\,
 e^{-i\omega(t-t')}\Bigg\{
 \frac{j_0(\omega r)-\omega\beta\eta_0(\omega r)}
 {\sqrt{1+\omega^2\beta^2}}\\
 &\qquad\times
 \frac{j_0(\omega r')-\omega\beta\eta_0(\omega r')}
 {\sqrt{1+\omega^2\beta^2}}
 -j_0(\omega r)j_0(\omega r')\Bigg\}.
\end{aligned}
\end{equation}
Using the orthogonality of the spherical harmonics,
\begin{equation}
    \int d\Omega_k \, Y_\ell^m(\theta_k,\varphi_k)Y_{\bar{\ell}}^{*\bar{m}}(\theta_k,\varphi_k)
    = \delta_{\ell, \bar{\ell}} \,\delta_{m,\bar{m}},
\end{equation}
together with the identity
\begin{equation}
    e^{i\, \mathbf{k}\cdot\mathbf{r}} =
    4\pi\sum_{\ell,m}i^\ell j_\ell(k r)Y_\ell^m(\theta,\phi)Y_{\ell}^{*m}(\theta_k,\phi_k),
\end{equation}
it is straightforward to show that
\begin{equation}\label{wightmandirichlet}
    W_{\mathrm M}(x,x') = \dfrac{1}{(2\pi)^3}\int\dfrac{d^3k}{2\omega}\, e^{-i\,k\cdot (x-x')},
\end{equation}
i.e., \(W_{\mathrm M}(x,x')\) is the usual Wightman function in Minkowski spacetime.

\section{Accelerated Detector Response}\label{sec:detector-response}

\subsection{Detector model and trajectories}\label{subsec:detector-setup}

At leading order in the coupling, the response of a two-level UDW detector with energy gap \(\Omega=E-E_0\) is
\begin{equation}\label{response}
F(\Omega)
=
\lim_{\epsilon\to0^+}
\int_{-\infty}^{\infty}d\tau
\int_{-\infty}^{\infty}d\tau'\,
\chi(\tau)\chi(\tau')
 e^{-i\Omega(\tau-\tau')}
 W_\epsilon(\tau,\tau'),
\end{equation}
where \(\chi\) is a real switching function and
\(W_\epsilon(\tau,\tau')=W_\epsilon(x(\tau),x(\tau'))\) is the regularized Wightman function pulled back to the detector worldline. If the pullback is stationary, \(W_\epsilon(\tau,\tau')=W_\epsilon(\tau-\tau')\), one may define the time-independent transition rate
\begin{equation}\label{rate}
\dot F(\Omega)=
\lim_{\epsilon\to0^+}
\int_{-\infty}^{\infty}ds\,
 e^{-i\Omega s}W_\epsilon(s).
\end{equation}

For an accelerated detector in the Minkowski vacuum, we  focus first on the unshifted, radially aligned trajectory
\begin{equation}\label{trajectory}
x^\mu(\tau)=
\left(
 a^{-1}\sinh(a\tau),
 a^{-1}\cosh(a\tau),
 0,
 0
\right).
\end{equation}
Its radial distance from the puncture is \(r(\tau)=a^{-1}\cosh(a\tau)\), so it reaches the minimum distance \(r_{\min}=1/a\) at \(\tau=0\) and then recedes.

For the ordinary Minkowski contribution, substituting Eq.~\eqref{trajectory} into Eq.~\eqref{rate} gives
\begin{equation}
\dot F_{\mathrm M}(\Omega)=
\frac{1}{2\pi}\frac{\Omega}{e^{2\pi\Omega/a}-1},
\end{equation}
which satisfies the detailed-balance relation~\cite{martinez}
\begin{equation}
\frac{\dot F_{\mathrm M}(-\Omega)}{\dot F_{\mathrm M}(\Omega)}
=e^{2\pi\Omega/a}
\end{equation}
at the Unruh temperature \(T_{\mathrm U}=a/(2\pi)\).

Because Eq.~\eqref{response} is linear in the two-point function, the decomposition
\(W=W_{\mathrm M}+W_\beta\) induces
\(F=F_{\mathrm M}+F_\beta\). We call \(F_\beta\) the boundary-induced, subtracted response. It is a difference of responses rather than a transition probability by itself and therefore need not be positive.

\subsection{Boundary-induced Wightman function}\label{subsec:boundary-wightman}
We first subtract the Minkowski term \(W_{\mathrm M}(x,x')\) given by Eq.~\eqref{wightmandirichlet} from the full Wightman function. For \(\beta>0\), define
\begin{equation}
 R:=r+r',\qquad \Delta t:=t-t',\qquad
 w:=R+\Delta t,\qquad v:=R-\Delta t.
\end{equation}
Using \(j_0(z)=\sin z/z\) and \(\eta_0(z)=-\cos z/z\), Eq.~\eqref{wightmanrobinexpansion} reduces to
\begin{equation}
\begin{aligned}
 W_\beta(x,x')&=
 \frac{1}{4\pi^2rr'}\lim_{\epsilon\to0^+}
 \int_0^\infty d\omega\,
 e^{-i\omega(\Delta t-i\epsilon)}\\
 &\qquad\times
 \frac{\beta\sin(\omega R)+\beta^2\omega\cos(\omega R)}
 {1+\beta^2\omega^2}.
\end{aligned}
 \label{wightmanrobin-reduced}
\end{equation}
Retaining the Wightman prescription \(\Delta t\to\Delta t-i0\), this integral can be written globally using Heaviside functions as~\cite{gradshteyn}
\begin{widetext}
\begin{align}
 W_\beta(x,x')=\frac{1}{8\pi^2rr'}\Bigg\{&
 e^{w/\beta}\left[
 \Theta(w)E_1\!\left(\frac{w}{\beta}\right)
 -\Theta(-w)\operatorname{Ei}\!\left(-\frac{w}{\beta}\right)
 +i\pi\Theta(-w)
 \right]\nonumber\\
 &+e^{v/\beta}\left[
 \Theta(v)E_1\!\left(\frac{v}{\beta}\right)
 -\Theta(-v)\operatorname{Ei}\!\left(-\frac{v}{\beta}\right)
 -i\pi\Theta(-v)
 \right]\Bigg\}.
 \label{wightmanrobin-global}
\end{align}
\end{widetext}
Here \(E_1(z)=\int_z^\infty e^{-q}\,dq/q\) for \(z>0\), and \(\operatorname{Ei}\) is the exponential integral on the real axis. Equation~\eqref{wightmanrobin-global} is equivalently the boundary value
\begin{equation}
 W_\beta(x,x')=
 \frac{e^{w/\beta}E_1[(w-i0)/\beta]
 +e^{v/\beta}E_1[(v+i0)/\beta]}
 {8\pi^2rr'},
 \label{wightmanrobin-i0}
\end{equation}
with the branch cut of \(E_1\) along the negative real axis. The step-function form makes the two boundary values across the branch cut explicit.

For the uniformly accelerated trajectory~\eqref{trajectory}, one has
\begin{equation}
 r(\tau)+t(\tau)=a^{-1}e^{a\tau},
 \qquad
 r(\tau)-t(\tau)=a^{-1}e^{-a\tau},
\end{equation}
so that
\begin{align}
 w(\tau,\tau')&:=r(\tau)+r(\tau')+t(\tau)-t(\tau')
 =a^{-1}\left(e^{a\tau}+e^{-a\tau'}\right)>0,\nonumber\\
 v(\tau,\tau')&:=r(\tau)+r(\tau')-t(\tau)+t(\tau')
 =a^{-1}\left(e^{-a\tau}+e^{a\tau'}\right)>0.
\end{align}
All Heaviside factors in Eq.~\eqref{wightmanrobin-global} therefore reduce unambiguously, and the explicitly imaginary terms vanish. The pullback \(W_\beta(\tau,\tau'):=W_\beta(x(\tau),x(\tau'))\) is
\begin{equation}\label{wightmanrobin}
 W_\beta(\tau,\tau')=
 \frac{1}{8\pi^2 r(\tau)r(\tau')}
 \left[
 e^{w/\beta}E_1\!\left(\frac{w}{\beta}\right)
 +e^{v/\beta}E_1\!\left(\frac{v}{\beta}\right)
 \right].
\end{equation}
The Dirichlet case \(\beta=0\) is recovered continuously from the large-argument asymptotic behavior \(e^zE_1(z)=z^{-1}+O(z^{-2})\), which gives \(W_\beta\to0\) as \(\beta\to0^+\). Moreover, for finite \(\beta>0\), the coincidence limit ($r^\prime\to r$ and $t^\prime\to t$) is
\begin{equation}
 W_\beta(r)=
 \frac{e^{2r/\beta}E_1\!\left(2r/\beta\right)}
 {4\pi^2r^2},
\end{equation}
which is finite. Thus the usual local coincidence singularity of the full Wightman function is entirely contained in \(W_{\mathrm M}\).  Moreover, equation~\eqref{wightmanrobin} is manifestly symmetric under \(\tau\leftrightarrow\tau'\), which interchanges \(w\) and \(v\). Hence
\begin{equation}
 W_\beta(\tau,\tau')=W_\beta(\tau',\tau)\in\mathbb{R}.
 \label{wbeta-symmetry}
\end{equation}

\subsection{Proper-time nonstationarity and comparison with thermal states}\label{subsec:stationarity}

The static ground state \(|0_\beta\rangle\) is invariant under simultaneous inertial-time translations of its arguments. Proper-time translations along the trajectory~\eqref{trajectory}, however, are generated by Lorentz boosts in the \((t,x)\) plane. In ordinary Minkowski spacetime, Poincar\'e invariance implies that the pullback of \(W_{\mathrm M}\) is stationary under this flow, and its analytic continuation satisfies the KMS condition at inverse temperature \(2\pi/a\). The standard time-independent Unruh transition rate then follows.

The timelike line removed from the spacetime, \(\mathcal L=\{(t,\mathbf 0):t\in \mathbb{R}\}\), is not invariant under these boosts. The Robin extension therefore preserves the static time translations generated by \(\partial_t\) but not the boost flow followed by the detector. Consequently, in general,
\begin{equation}
 W_\beta(\tau+s,\tau'+s)\neq W_\beta(\tau,\tau'),
 \label{nonstationary}
\end{equation}
and the boundary-induced pullback is not stationary in the detector's proper time. Since stationarity is a prerequisite for a KMS condition with respect to that evolution, the full pullback cannot be assigned a proper-time KMS temperature, and the boundary-induced term does not define a time-independent transition rate. 

\begin{figure}[tbp]    \centering\includegraphics[width=.48\textwidth]{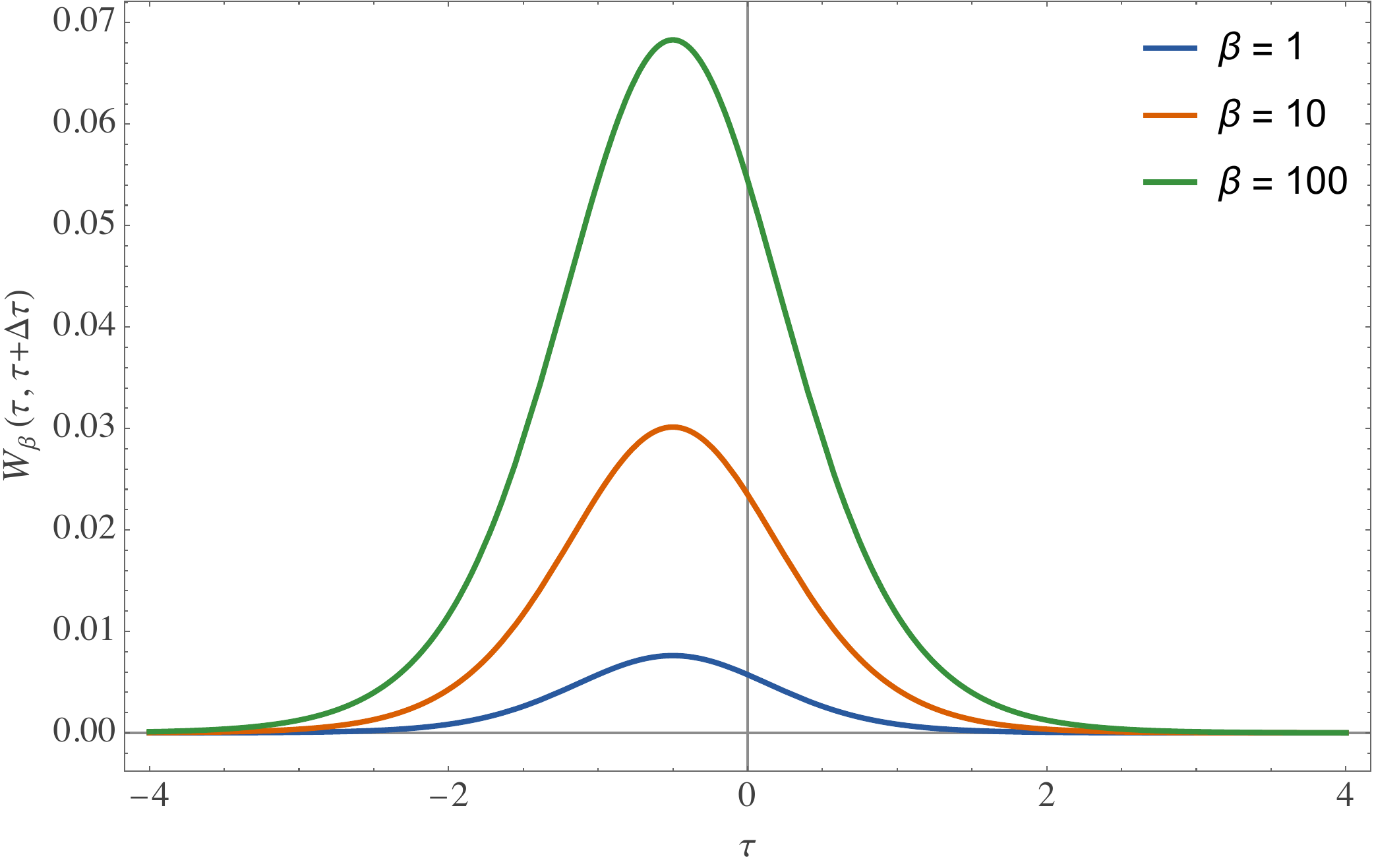}
    \caption{\(W_\beta(\tau,\tau+\Delta\tau)\) for the uniformly accelerated trajectory~\eqref{trajectory} as a function of \(\tau\), with the fixed offset \(\Delta\tau=1\) and acceleration \(a=1\). If \(W_\beta\) depended only on the proper-time separation, it would be constant in \(\tau\) for every \(\beta\). The variation shown here demonstrates the failure of proper-time stationarity in Eq.~\eqref{nonstationary}.}
    \label{fig:wightman_robin_delta_tau}
\end{figure}

This also clarifies the distinction from Ref.~\cite{pitelli1}. The finite-temperature detector analyzed there is static and probes a thermal KMS state with respect to the same inertial-time translations preserved by the boundary condition, so its pullback remains stationary. Here the detector accelerates through the static ground state, and the puncture breaks the boost symmetry that, in complete Minkowski spacetime, makes the vacuum pullback along a uniformly accelerated orbit a KMS correlation function. Fig.~\ref{fig:wightman_robin_delta_tau} directly illustrates the resulting proper-time nonstationarity.

Having found the pullback of the Wightman function, we evaluate the response in Eq.~\eqref{response} along the trajectory~\eqref{trajectory}. With \(s=\tau-\tau'\) and \(u=\tau\), the boundary-induced contribution can be written as
\begin{equation}
 F_\beta(\Omega)=2\int_{-\infty}^{\infty}du\,\chi(u)
 \int_0^{\infty}ds\,
 \mathrm{Re}\!\left[e^{-i\Omega s}\chi(u-s)W_\beta(u,u-s)\right].
 \label{fbeta}
\end{equation}
Because the switching function is real and Eq.~\eqref{wbeta-symmetry} holds, one may equivalently write
\begin{equation}
 F_\beta(\Omega)=
 \int_{-\infty}^{\infty}d\tau\int_{-\infty}^{\infty}d\tau'\,
 \chi(\tau)\chi(\tau')
 \cos\!\left[\Omega(\tau-\tau')\right]W_\beta(\tau,\tau'),
\end{equation}
and therefore
\begin{equation}
 F_\beta(-\Omega)=F_\beta(\Omega).
\label{fbeta-even}
\end{equation}
To avoid the ultraviolet divergences associated with abruptly switching a pointlike detector on and off~\cite{matsas,satz1}, and to keep the comparison with the Minkowski response well defined, we use a smooth switching function. Specifically, we choose a profile that approximates a rectangular window while switching the interaction on and off smoothly, namely
\begin{align}\label{switchingfunctions}
     \chi_k(\tau,\tau_0,T) =&\, \Theta_k(\tau-\tau_0) - \Theta_k(\tau-\tau_0-T), \nonumber \\
     \Theta_k(\tau) =& \dfrac{1}{2} \left[1+\tanh(k\tau)\right].
\end{align}

\subsection{Long-time behavior}\label{subsec:infinite-time}

We now establish the long-time behavior of the boundary-induced response along the trajectory~\eqref{trajectory}. To extend the detector-field interaction over increasingly long portions of the same worldline, we set \(\tau_0=-T/2\) in Eq.~\eqref{switchingfunctions}, so that the smooth switching profile expands symmetrically about the instant of closest approach:
\begin{equation}\label{long-time-window}
 \chi_T(\tau):=\chi_k(\tau,-T/2,T)
 =\Theta_k(\tau+T/2)-\Theta_k(\tau-T/2).
\end{equation}
Then \(0\leq\chi_T(\tau)\leq1\) and, for every fixed \(\tau\), \(\chi_T(\tau)\to1\) as \(T\to\infty\).

For \(z>0\), let \(H(z):=e^zE_1(z)\). A change of integration variable in the definition of \(E_1\) gives
\begin{equation}\label{H-bound}
 H(z)=\int_0^\infty dp\,\frac{e^{-p}}{z+p}\Longrightarrow
 0<H(z)\leq\frac{1}{z}.
\end{equation}
Because $w,v>0$ on this worldline, Eq.~\eqref{wightmanrobin} defines a continuous, real, and nonnegative function of $(\tau,\tau')$. For \(0<\beta<\infty\), Eq.~\eqref{H-bound} then gives
\begin{equation}\label{Wpointwise-bound}
 0\leq W_\beta(\tau,\tau')
 \leq
 \frac{\beta}{8\pi^2r(\tau)r(\tau')}
 \left(\frac{1}{w}+\frac{1}{v}\right).
\end{equation}
Introduce the dimensionless variables
\begin{equation}
 \xi=\frac{a}{2}(\tau+\tau'),
 \qquad
 \eta=\frac{a}{2}(\tau-\tau').
\end{equation}
Along the trajectory~\eqref{trajectory}, one has
\begin{align}
 r(\tau)r(\tau')
 &=\frac{1}{a^2}\left(\cosh^2\xi+\sinh^2\eta\right),\nonumber\\
 w&=\frac{2}{a}e^\eta\cosh\xi,
 \qquad
 v=\frac{2}{a}e^{-\eta}\cosh\xi,\nonumber\\
 d\tau\,d\tau'&=\frac{2}{a^2}\,d\xi\,d\eta.
\end{align}
Therefore,
\begin{align}
 &\int_{\mathbb R^2}d\tau\,d\tau'\,
 W_\beta(\tau,\tau')\nonumber\\
 &\quad\leq
 \frac{a\beta}{4\pi^2}
 \int_{-\infty}^{\infty}d\xi
 \int_{-\infty}^{\infty}d\eta\,
 \frac{\cosh\eta}
 {\cosh\xi\left(\cosh^2\xi+\sinh^2\eta\right)}.
 \label{W-L1-intermediate}
\end{align}
For fixed \(\xi\), we have
\begin{equation}
 \int_{-\infty}^{\infty}d\eta\,
 \frac{\cosh\eta}
 {\cosh\xi\left(\cosh^2\xi+\sinh^2\eta\right)}
 =\frac{\pi}{\cosh^2\xi}.
\end{equation}
Consequently,
\begin{equation}\label{W-L1}
 \int_{\mathbb R^2}d\tau\,d\tau'\,
 W_\beta(\tau,\tau')
 \leq \frac{a\beta}{2\pi}<\infty.
\end{equation}
Thus the boundary-induced pullback is absolutely integrable for every fixed finite \(\beta>0\); the Dirichlet case \(\beta=0\) is trivial because \(W_0=0\).

For the switching profile~\eqref{long-time-window}, write the response as
\begin{equation}\label{FbetaT}
 F_{\beta,T}(\Omega)=
 \int_{\mathbb R^2}d\tau\,d\tau'\,
 \chi_T(\tau)\chi_T(\tau')
 \cos\!\left[\Omega(\tau-\tau')\right]
 W_\beta(\tau,\tau').
\end{equation}
The absolute value of the integrand is bounded by the integrable function \(W_\beta(\tau,\tau')\). Hence the dominated convergence theorem gives
\begin{align}
 \lim_{T\to\infty}F_{\beta,T}(\Omega)
 &=F_{\beta,\infty}(\Omega)\nonumber\\
 &:=%
 \int_{\mathbb R^2}d\tau\,d\tau'\,
 \cos\!\left[\Omega(\tau-\tau')\right]
 W_\beta(\tau,\tau'),
 \label{Fbeta-limit}
\end{align}
where the limiting integral is absolutely convergent. Moreover,
\begin{equation}\label{Fbeta-uniform-bound}
 \left|F_{\beta,T}(\Omega)\right|
 \leq\frac{a\beta}{2\pi}
\end{equation}
for all \(T\), \(\Omega\), and fixed finite \(\beta\). In particular, \(F_{\beta,T}(\Omega)=O(1)\) and \(F_{\beta,T}(\Omega)/T\to0\): the puncture does not generate an additional term linear in the interaction duration. The estimate~\eqref{Fbeta-uniform-bound} grows with \(\beta\) and is therefore not uniform in the formal Neumann limit, consistently with the behavior discussed below.

\subsection{Neumann limit}\label{subsec:neumann}

The formal Neumann limit is singular. For $z\to0^+$,
\begin{equation}
 e^zE_1(z)=-\gamma-\ln z+O\!\left(z|\ln z|\right).
\end{equation}
Since $w,v>0$ on the accelerated trajectory, Eq.~\eqref{wightmanrobin} gives the dimensionally explicit pointwise asymptotics
\begin{equation}\label{wbeta-neumann}
 W_\beta(\tau,\tau')=
 \frac{\ln\!\left(\beta^2/(wv)\right)-2\gamma}
 {8\pi^2r(\tau)r(\tau')}+o(1),
 \qquad \beta\to\infty,
\end{equation}
for fixed $\tau$ and $\tau'$. On the trajectory~\eqref{trajectory}, $wv=4a^{-2}\cosh^2\xi$, so the natural dimensionless large parameter is $a\beta$. Equation~\eqref{wbeta-neumann} establishes that the boundary-induced two-point function has no finite Neumann limit and identifies the divergence as an infrared effect of the $s$-wave sector, rather than an ultraviolet divergence~\cite{dappiaggi}. For the switching profiles used in our numerical analysis, the detector response exhibits growth consistent with this logarithmic behavior.

\subsection{Numerical illustrations}\label{subsec:numerics}

We evaluate Eq.~\eqref{fbeta} numerically in Mathematica. In all numerical plots we set $a=1$ and $k/a=10$, so that $k=10$ in the units used in Eq.~\eqref{switchingfunctions}. The displayed quantities therefore represent the dimensionless combinations $\Omega/a$, $aT$, $a\beta$, and $k/a$.

\begin{figure*}[tbp]
    \centering
    \begin{minipage}[t]{0.48\textwidth}
        \centering
        \textbf{(a)}\\
        \includegraphics[width=\linewidth]{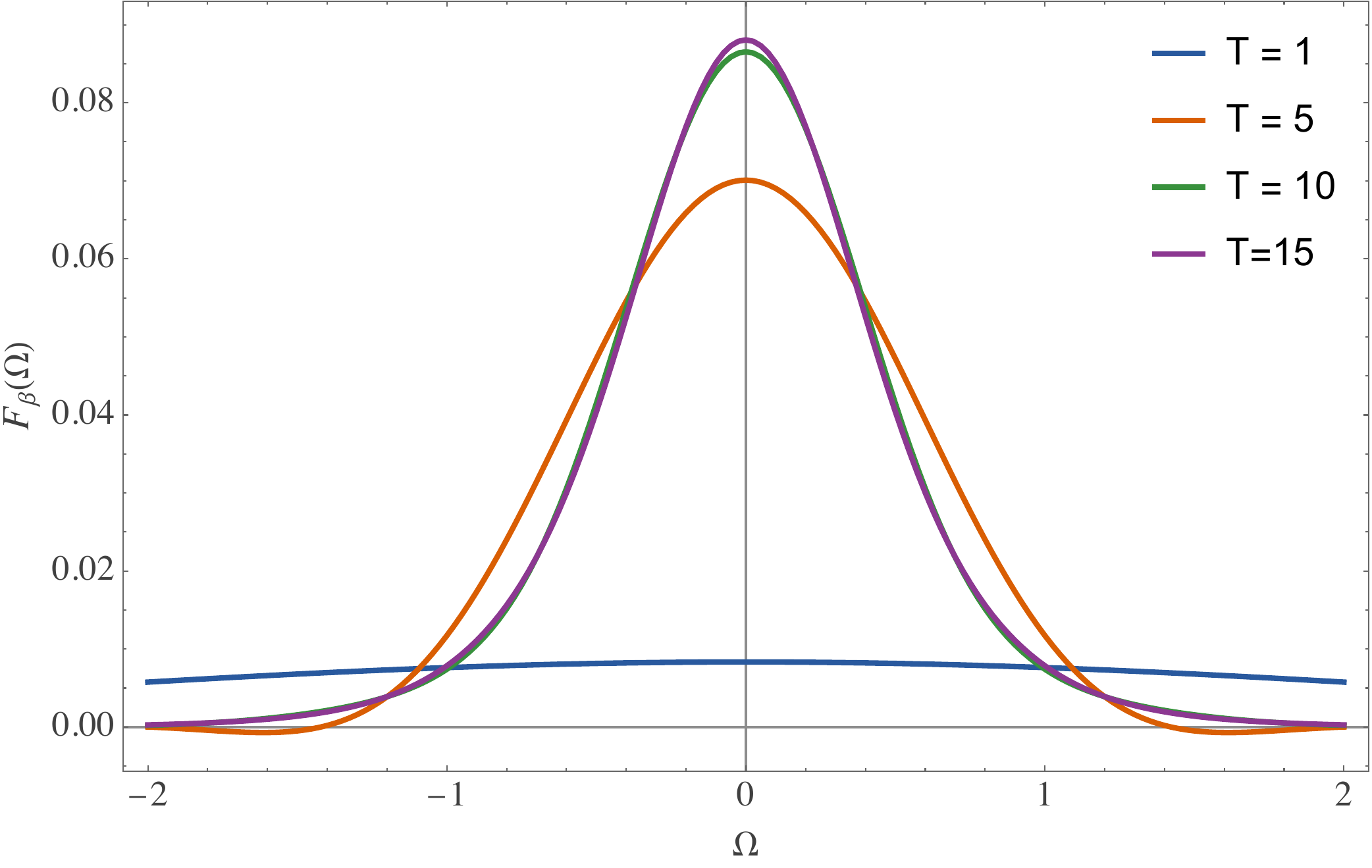}
    \end{minipage}\hfill
    \begin{minipage}[t]{0.48\textwidth}
        \centering
        \textbf{(b)}\\
        \includegraphics[width=\linewidth]{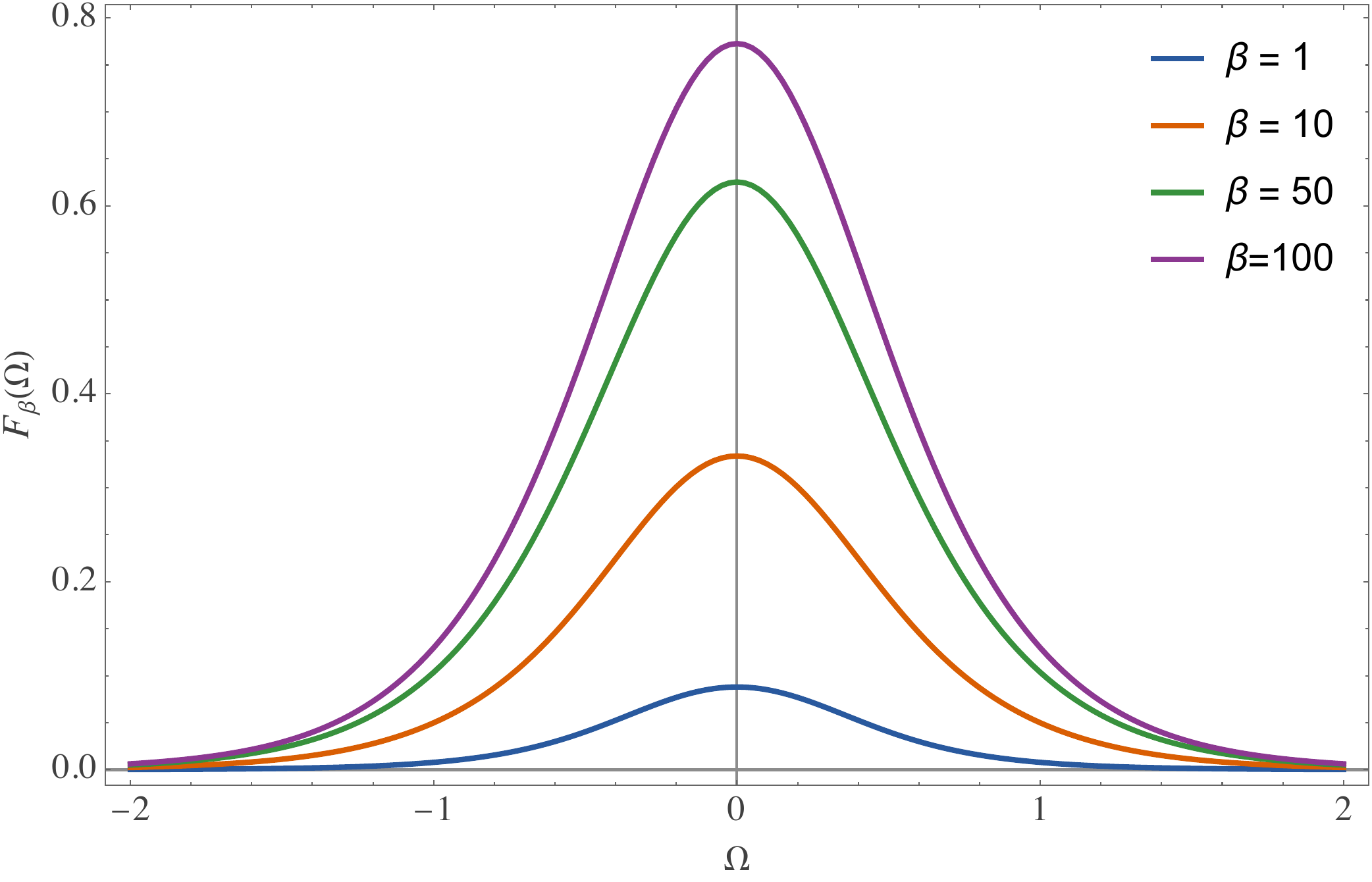}
    \end{minipage}
    \caption{Numerical response profiles. (a) $F_\beta(\Omega)$ for $\beta=1$ and $\tau_0=-T/2$, shown for several interaction durations; the near superposition of the $T=10$ and $T=15$ curves illustrates the approach to the finite large-$T$ limit. (b) $F_\beta(\Omega)$ for $\tau_0=-T/2$ and $T=15$, shown for several values of $\beta$; the magnitude changes as the formal Neumann limit is approached.}
    \label{fig:numerics}
\end{figure*}

Figure~\ref{fig:numerics}(a) illustrates the finite-duration response for $\beta=1$ with $\tau_0=-T/2$. The largest-$T$ curves are nearly superposed, as expected from the theorem in Sec.~\ref{subsec:infinite-time}. Equation~\eqref{fbeta-even} explains their exact symmetry under $\Omega\to-\Omega$. Depending on the gap and duration, the subtracted contribution can have either sign: a negative value suppresses the corresponding Minkowski response, whereas a positive value enhances it. At fixed large duration, the magnitude and shape depend strongly on the Robin parameter, as shown in Fig.~\ref{fig:numerics}(b). The increasing magnitude as the formal Neumann limit is approached is consistent with the logarithmic infrared behavior in Eq.~\eqref{wbeta-neumann}.

\begin{figure}[!htb]
    \centering \hspace{-.33cm}\includegraphics[width=.5\textwidth]{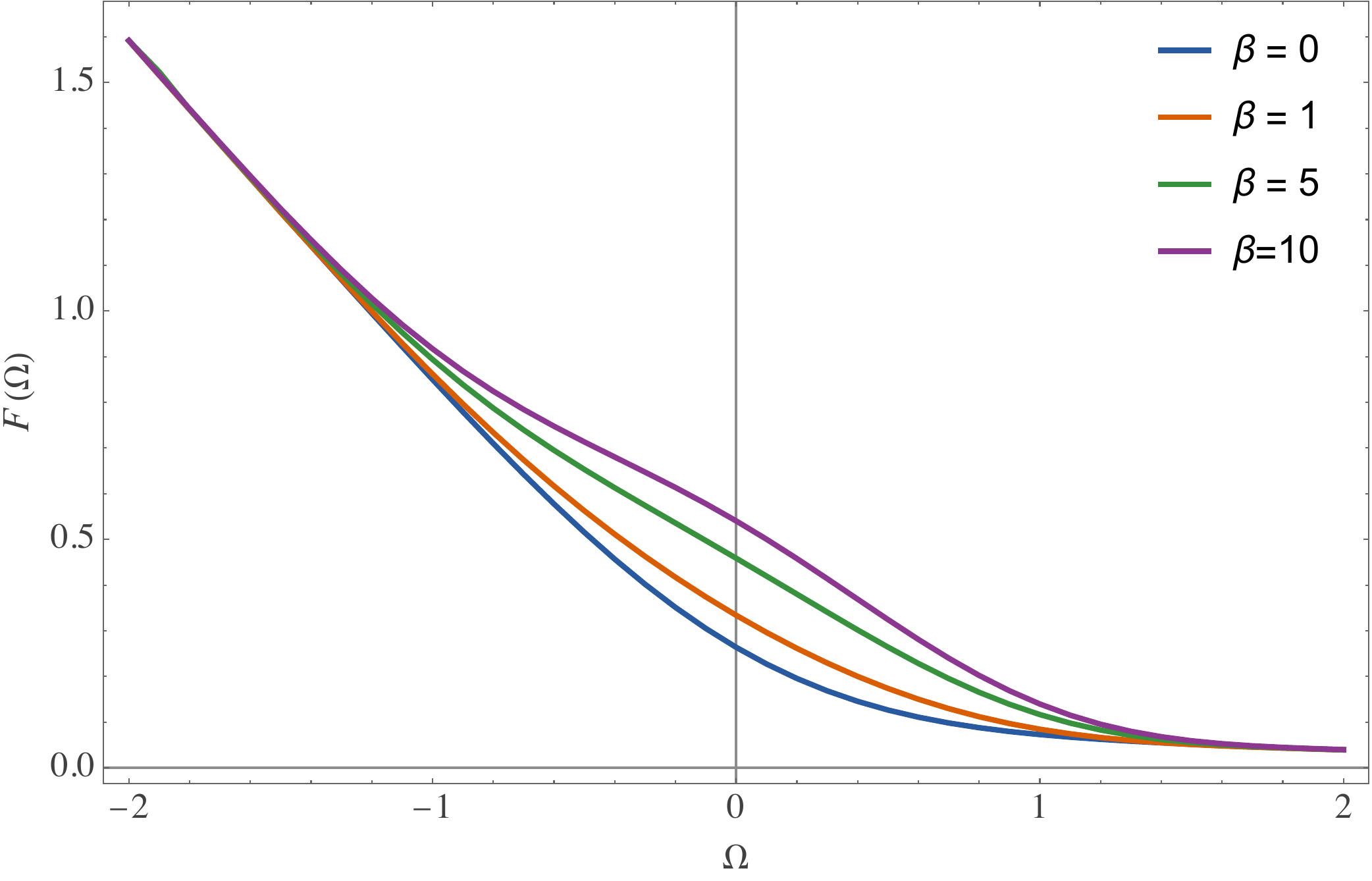}
    \caption{Full response function \(F(\Omega)=F_\beta(\Omega)+F_{\mathrm M}(\Omega)\), shown for \(\tau_0=-T/2\), \(T=5\), and different values of the parameter \(\beta\). It remains positive for all cases considered, even when the subtracted contribution \(F_\beta(\Omega)\) is negative. This is consistent with the interpretation of the full response \(F(\Omega)\), not of \(F_\beta\) separately, as a quantity proportional to the transition probability.} 
    \label{fig:robin_completa}
\end{figure}

The fact that $F_\beta$ can be negative is not problematic: it is a difference of responses. The full response $F=F_{\mathrm M}+F_\beta$ is nonnegative for every real switching function because it is obtained by smearing the positive-type Wightman function of the static ground state. The sign of the subtracted term therefore measures enhancement or suppression relative to the Minkowski response, not the positivity of a standalone probability. Figure~\ref{fig:robin_completa} shows that this is indeed what happens in the cases considered: the full response remains strictly positive, as expected for a quantity proportional to a transition probability at leading perturbative order.

\section{Dependence on the Trajectory and Interaction Window}\label{sec:distance}

The puncture singles out the spatial origin, making the detector response sensitive to the placement of the accelerated trajectory. We probe this dependence by translating the hyperbola longitudinally and by varying the portion of the worldline sampled by the detector.

\begin{figure}[tbp]
    \centering\includegraphics[width=.48\textwidth]{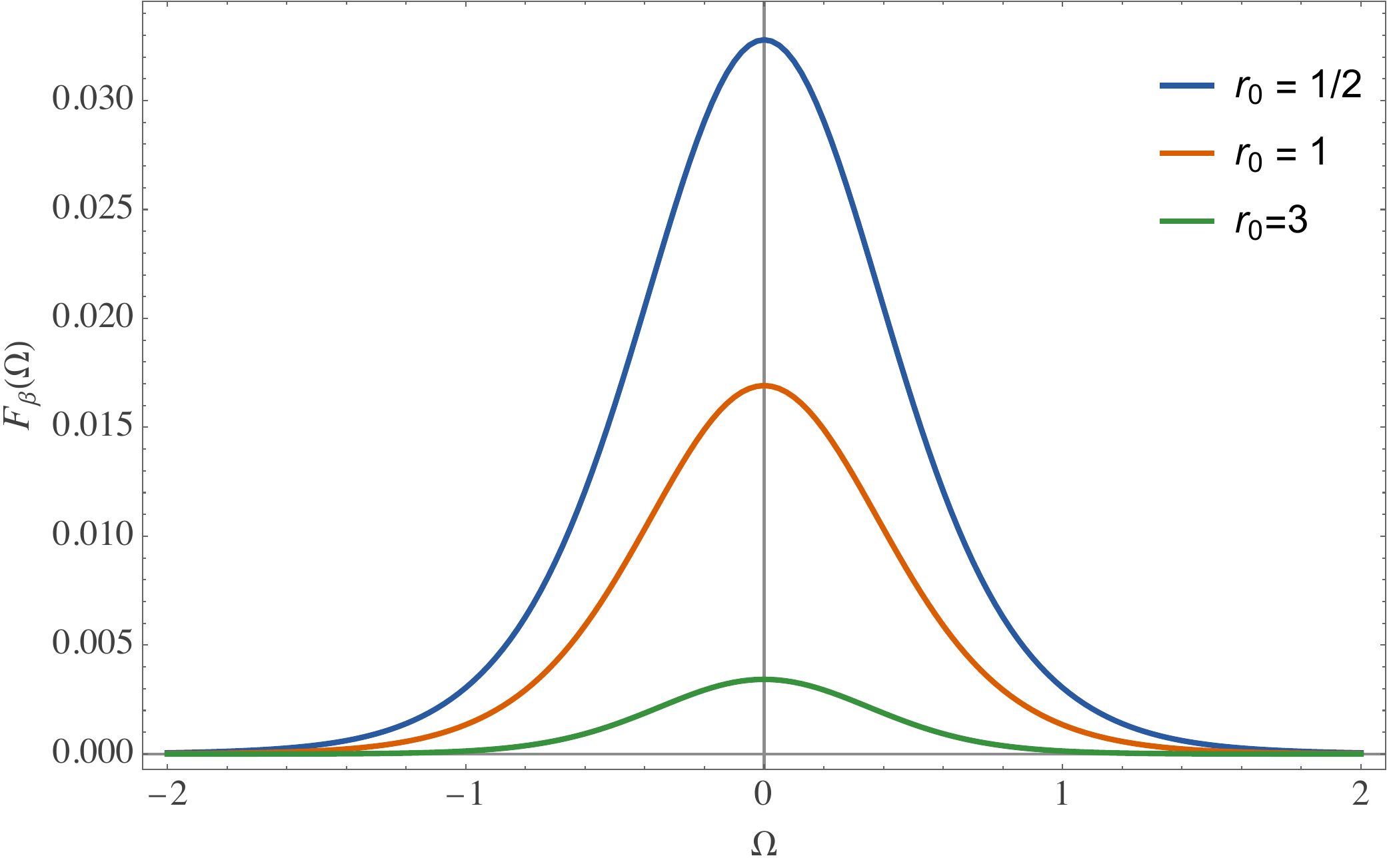}
    \caption{\(F_\beta(\Omega)\) for a uniformly accelerated trajectory rigidly displaced by \(r_0\), with \(\beta=10\), \(\tau_0=-T/2\), and \(T=15\), for different values of the displacement parameter \(r_0\). Increasing \(r_0\) reduces the magnitude of the boundary-induced effect.}
    \label{fig:robin_r_desloc_b10}
\end{figure}

We first probe the longitudinal displacement by replacing \(x(\tau)=a^{-1}\cosh(a\tau)\) with \(x_{r_0}(\tau)=a^{-1}\cosh(a\tau)+r_0\) while leaving \(t(\tau)\) unchanged. This translation leaves the detector's proper acceleration unchanged and, for the positive values used below, increases the minimum distance to \(r_{\min}=a^{-1}+r_0\). Figure~\ref{fig:robin_r_desloc_b10} shows that the magnitude of the boundary-induced response decreases accordingly. 

\begin{figure}[tbp]
    \centering\includegraphics[width=.48\textwidth]{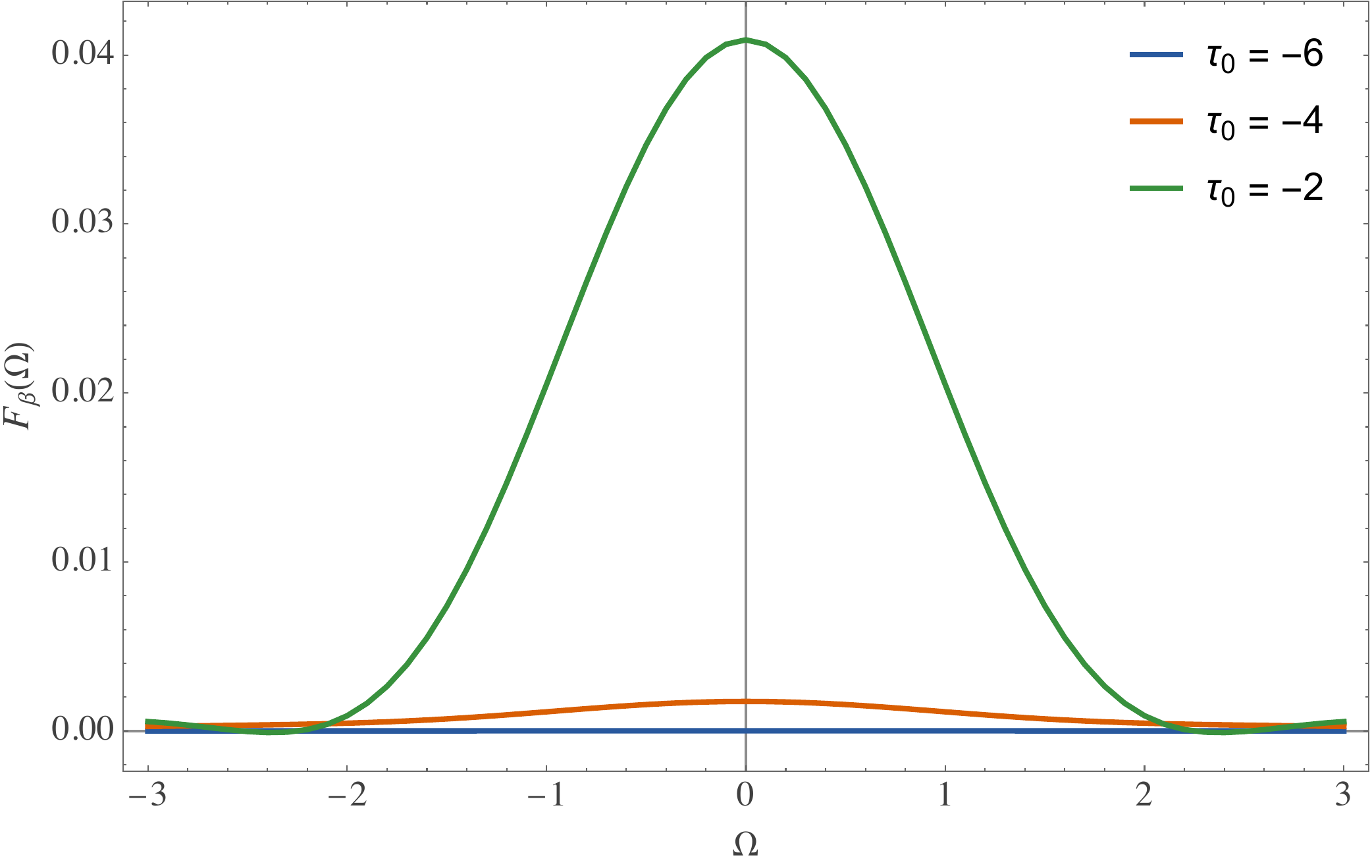}
    \caption{\(F_\beta(\Omega)\) for different interaction windows, with \(\beta=10\), \(T=3\), and varying activation times \(\tau_0\). The boundary-induced effect is almost negligible far from the puncture, as in the case \(\tau_0=-6\), but becomes increasingly pronounced as the detector approaches the puncture.}
    \label{fig:robin_switching_desloc_b10}
\end{figure}

Finally, we keep the trajectory given in Eq.~\eqref{trajectory}, but change the switching function so that the detector interacts with the field during different portions of the trajectory. Rather than centering the smooth switching profile around \(\tau=0\), as in the choice \(\tau_0=-T/2\), we fix \(T=3\) and vary \(\tau_0\) in order to probe the boundary-induced effect at different stages of the detector's motion. Figure~\ref{fig:robin_switching_desloc_b10} shows that the response becomes more pronounced as the switching window moves toward the interval of closest approach.

\section{Conclusion}\label{sec:conclusion}

We studied a real massless minimally coupled scalar field on punctured Minkowski spacetime with the stable Robin extensions \(G(0)-\beta G'(0)=0\), $\beta\geq 0$, choosing for each finite \(\beta\) the static ground state defined by the inertial Killing time. 

The Wightman function splits as \(W=W_{\mathrm M}+W_\beta\). We first obtained a closed form for \(W_\beta\). For the radially aligned trajectory, the pullback of $W_\beta$ is real and symmetric but nonstationary because the puncture preserves inertial-time translations while breaking the relevant boost flow. The ordinary Minkowski term retains the Unruh KMS property, whereas the boundary-induced term has neither a proper-time KMS temperature nor a time-independent transition rate. The corresponding subtracted response is even in the detector gap and can enhance or suppress the Minkowski response.

Our main analytic result is that, for every fixed finite \(\beta\), this pullback is absolutely integrable in the two proper-time variables. As the interaction is extended over the full detector history, the response has a finite \(T\to\infty\) limit and remains \(O(1)\); it therefore generates no additional term linear in the interaction duration. The formal Neumann limit is qualitatively different: the boundary-induced two-point function diverges logarithmically because of an infrared singularity in the \(s\)-wave sector, and the numerical detector response exhibits growth consistent with this asymptotic behavior.

The numerical illustrations display the approach to the long-time limit and demonstrate that the correction depends strongly on \(\beta\) and on the detector's location relative to the puncture. We analyzed the unshifted, radially aligned hyperbola and a longitudinally displaced slice of a broader family of inequivalent accelerated trajectories. Punctured Minkowski spacetime thus provides an exactly tractable setting in which boundary conditions, state choice, broken boost symmetry, and detector observables can be separated cleanly; it also connects directly with the zero-solid-angle-deficit limit of the idealized global-monopole problem when the origin is kept removed.

\begin{acknowledgments}
N.P.B. acknowledges financial support from Coordena\c{c}\~ao de Aperfei\c{c}oamento de Pessoal de N\'ivel Superior (CAPES, Brazil) under Grant No.~88887.947715/2024-00.  J.P.M.P. acknowledges partial support from Conselho Nacional de Desenvolvimento
Cient\'ifico e Tecnol\'ogico (CNPq, Brazil) under Grant No.~305194/2025-9 and
Funda\c{c}\~ao de Amparo a Pesquisa do Estado de S\~ao
Paulo (FAPESP), Grant No. 2024/00923-6.
R.A.M. was partially supported by Conselho Nacional de Desenvolvimento
Cient\'ifico e Tecnol\'ogico (CNPq, Brazil) under Grant No.~316780/2023-5 and
 Funda\c{c}\~ao de Amparo a Pesquisa do Estado de S\~ao
Paulo (FAPESP), Grant No. 2024/00923-6.

The authors thank the anonymous referee for a careful and constructive report,
which led to substantial improvements in the manuscript.

The authors used OpenAI ChatGPT (GPT-5.6) and OpenAI Codex tools as interactive
aids for exploratory discussion, editing, LaTeX preparation, and auxiliary
code/plot checks.  The authors remain fully responsible for this work.
\end{acknowledgments}


\end{document}